# A Generic Solution of Fermion Sign Problem

J. Wang[a] and D. Y. Sun[a,b*]
[a]*Department of Physics, East China Normal University, 200241 Shanghai, China*
[b]*Shanghai Qi Zhi Institute, Shanghai 200030, China*

## Abstract

The fermion sign problem, the biggest obstacle in quantum Monte Carlo calculations, is completely solved in this paper. Here, we find a strategy, in which the contribution from those negative-weighted paths is thoroughly cancelled or replaced by some positive-weighted paths. The crucial point lies on the Feynman path integral formula proposed in our group, which allows us to deeply analyze the Boltzmann weight of each path. Through mathematical proof, we demonstrate that physical quantities can be exactly calculated within a specific kind of paths, which have positive the Boltzmann weight. With this finding, a new Monte Carlo method is proposed, in which the fermion sign problem is absent. As an example, the current method is applied to the two-dimensional Hubbard model, and the results do manifest the correctness.

[*]Corresponding author: dysun@phy.ecnu.edu.cn

# I. Introduction

In quantum Monte Carlo (QMC) simulations, the notorious fermion sign problem (SP) has become a formidable obstacle for nonperturbative studies of various many-fermion systems, ranging from chemistry,[1] cold atoms,[2] condensed matter[3, 4] to particle and high-energy physics.[5, 6] Regarding to the SP, the computational physics seems to be in a helpless theoretical situation.[7, 8] For many years, albeit great efforts, the SP can only be relieved[9-46] or circumvented for some specific systems,[47-66] a generic solution is still missing. It is certain that, a complete solution of the SP will greatly promote the development of computational physics and many-fermion systems.

The SP has its roots in the antisymmetric character of many-fermion wave functions (WFs). The anti-symmetry makes most QMC calculations have to face a sum of numerically closed data but with the opposite sign or Boltzmann weight (BW), which results in the sign of data totally overwhelming QMC calculations. Far worse, the SP has the complicated dependence on systems,[67-70] the choice of basis,[57, 71-73] as well as algorithms.[74, 75] A recent study even shows that, the SP in determinant QMC could be quantitatively linked to quantum critical behavior.[7] Now, scientists have to accept the fact that, the SP is nondeterministic polynomial hard (NP-hard problem), and a generic solution is almost impossible within the known computational methods.[76]

Although the partition function of a many-fermion system is a sum of paths with both positive and negative BW, it must be a positive number. This implies that those negative-weighted paths must be completely cancelled out or replaced by some positive-weighted paths. If one can find out how those paths with negative BW are cancelled, the SP may be completely solved. Of course, this is not easy, because each path in the WF space is highly nonlocal and mutually entangled.

In this paper, based on the recently proposed formula of the path integral in field theory,[77] we are able to identify how these negative-weighted paths can be replaced. According to this finding, we suggest a new QMC framework, in which the SP is completely solved. We test our method on the two-dimensional Hubbard model, and the obtained results are in excellent with the previous ones.

# II. Mathematical Proof and Discussion

In this section, based on the strict mathematical derivation, we have proved that the physical properties of a many-fermion system can be exactly calculated within a specific kind of paths, in which the SP is absent. To illustrate the current method, we choose Hubbard model[78] as an example. This model has been systematically and deeply studied, which provides various computational data for comparing.[79-82] However, it should be stressed that our method can be straightforwardly extended to any model or Hamiltonian.

**General Formula:** The current work is based on the path integral formula recently proposed in our group.[77] Here, we first present a brief introduction to this formula. The key step in this formula is to combine each off-diagonal term in Hamiltonian and its Hermitian conjugate into pairs. These paired operators are Hermitian, which will be used in the further mathematic deduction. For the Hubbard model, the paired operators look like $h_{ij\sigma} = -t_{ij}\left(c_{i\sigma}^{\dagger}c_{j\sigma} + c_{j\sigma}^{\dagger}c_{i\sigma}\right)$, in which $t_{ij}$ is the hopping amplitude. In term of paired off-diagonal operators, the Hubbard Hamiltonian can be rewritten as:

$$H = \sum_{i,j,\sigma} h_{ij\sigma} + U\sum_{i} n_{i\uparrow}n_{i\downarrow}, \quad (1)$$

where all symbols have its standard meaning. For convenience, the spin index ($\sigma$) is omitted in the following context, it will be included later on to prevent confusion.

In the occupation number representation, let $|ijK\rangle$ denote a many-fermion WF, in which the occupancy of $i$-th and $j$-th sites is explicitly given and the occupancy of rest sites is simply labeled by $K$. For the $i$-th and $j$-th sites, the occupation can be classified into four cases, which are labeled as $|i\bar{j}K\rangle, |\bar{i}jK\rangle, |ijK\rangle,$ and $|\bar{i}\bar{j}K\rangle$. Here, $\bar{i}(\bar{j})$ indicates no electron occupying the $i$-th ($j$-th) site, while $i$ ($j$) indicates one electron occupying the $i$-th ($j$-th) site.

According to the previous work,[77] the non-zero matrix element of a local Boltzmann operator, $e^{-\tau h}$ ($h = h_{ij}$ or $Un_{i\uparrow}n_{i\downarrow}$, and $\tau$ being a real number), is only presented in following cases:

$$\langle i\bar{j}K|e^{-\tau h_{ij}}|\bar{i}jK'\rangle = \langle \bar{i}jK|e^{-\tau h_{ij}}|i\bar{j}K'\rangle = \frac{1}{2}\delta_{K,K'}(e^{\theta_{ij}\tau t_{ij}} - e^{-\theta_{ij}\tau t_{ij}}) \quad (2a)$$

$$\langle i\bar{j}K|e^{-\tau h_{ij}}|i\bar{j}K'\rangle = \langle \bar{i}jK|e^{-\tau h_{ij}}|\bar{i}jK'\rangle = \frac{1}{2}\delta_{K,K'}(e^{\theta_{ij}\tau t_{ij}} + e^{-\theta_{ij}\tau t_{ij}}) \quad (2b)$$

$$\langle ijK|e^{-\tau h_{ij}}|ijK'\rangle = \langle \bar{i}\bar{j}K|e^{-\tau h_{ij}}|\bar{i}\bar{j}K'\rangle = \delta_{K,K'} \quad (2c)$$

where $\theta_{ij}$ is the sign produced by particle exchange as $h_{ij}$ acting on $|i\bar{j}K\rangle$ or $|\bar{i}jK\rangle$. For the diagonal operator $e^{-\tau U n_{i\uparrow} n_{i\downarrow}}$, the only non-zero matrix element reads:

$$\langle i_{\uparrow} i_{\downarrow} K|e^{-\tau U n_{i\uparrow} n_{i\downarrow}}|i_{\uparrow} i_{\downarrow} K'\rangle = \delta_{K,K'} e^{-\tau U} \quad (2d)$$

The matrix element in Eq. 2b-2d is the diagonal scattering, which value is independent on the sign of $\theta_{ij}$ and always positive. The matrix element in Eq. 2a is positive (negative) if $\theta_{ij} = 1\ (-1)$ (note $t_{ij} > 0$ in the Hubbard model). For $\theta_{ij} = 1\ (-1)$, Eq. 2a represents the off-diagonal scattering with the positive (negative) local sign. It is worth noting that, $\theta_{ij}$ is not the sign of the whole path. The sign of a path is determined by the number of occurrences of matrix element of Eq. 2a in the path (see Ref.[77] for details). In the following, we will call $\theta_{ij}$ as the ***local sign***, and call the sign of a path as the ***global sign***.

According to the standard path integral formula and Suzuki-Trotter decomposition,[83] the partition function of the Hubbard model is,

$$Z = Tre^{-\beta H} = Tr \prod_{ij} e^{-\tau h_{ij}} \prod_i e^{-\tau U n_{i\uparrow} n_{i\downarrow}} \cdots \prod_{ij} e^{-\tau h_{ij}} \prod_i e^{-\tau U n_{i\uparrow} n_{i\downarrow}} \quad (3)$$

where the imaginary time or reciprocal temperature ($\beta = \frac{1}{k_B T}$) is divided into $M$ time-slices with “time step” $\tau = \beta/M$. To calculate the partition function, a complete set of states are inserted between each local Boltzmann operator ($e^{-\tau h},\ h = h_{ij}$ or $U n_{i\uparrow} n_{i\downarrow}$). If the matrix element of all $e^{-\tau h}$ in Eq.3 is non-zero, we call the corresponding WF sequence a **closed path** in WF spaces (or a **closed world line**). The partition function is denoted as $Z = \sum_{\omega} \rho(\omega)$, where $\rho(\omega)$ represents the BW of the closed path $\omega$.

According to our strategy mentioned above, we try to divide the closed paths into two classes, $\omega_+$ and $\omega_-$. Any path in $\omega_+$ has the positive global sign and only contains the positive local sign, *i.e,* $\theta_{ij}$ equals to $1$ in Eq. 2a-2d (Hereafter **OP** paths). All rest paths belong to $\omega_-$ (Hereafter **ON** paths). An ON path either contains the negative local sign or has the negative global sign. Evidently, ON paths may have the global positive or negative sign. Now the partition function can be written as,

$$Z = \sum_{\omega} \rho(\omega) = \sum_{\omega_+} \rho(\omega_+) + \sum_{\omega_-} \rho(\omega_-) = Z_+ + Z_-.$$

Obviously, if $\frac{Z_-}{Z_+}$ is a constant, the negative-weighted paths can be well cancelled or replaced, and the physical quantities can be exactly calculated within OP paths, accordingly the SP will be solved automatically. Our following proof is based on this idea.

Our proof starts from analyzing the contribution of $e^{-\tau h_{12}}$ in the 1st time slice (1ST) to energies. For that, we further write $Z_+ = Z_+^{(0)} + Z_+^{(1)} + Z_+^{(2)}$ and $Z_- = Z_-^{(0)} + Z_-^{(1)} + Z_-^{(2)}$, which is illustrated in Fig. 1. Here $Z_\pm^{(0)}$ only contains these paths, at which the matrix element of $e^{-\tau h_{12}}$ in 1ST takes Eq. 2c (black dot line in Fig. 1). $Z_\pm^{(1)}$ only contains these paths, at which the matrix element of $e^{-\tau h_{12}}$ in 1ST appears as Eq. 2b (red solid line in Fig. 1). $Z_\pm^{(2)}$ only contains these paths, at which the matrix element of $e^{-\tau h_{12}}$ in 1ST appears as Eq. 2a (green solid line in Fig. 1). It is worth noting that, $\theta_{12}$ in 1ST could be +1 or -1 in $Z_-^{(1)}$ and $Z_-^{(2)}$, but it can only take +1 in $Z_+^{(1)}$ and $Z_+^{(2)}$. Evidently, paths in $Z_+^{(0,1,2)}$ are always non-negative, while paths in $Z_-^{(0,1,2)}$ can be positive or negative. It needs to point out that, $Z_\pm^{(0)}$, $Z_\pm^{(1)}$ and $Z_\pm^{(2)}$ are classified regarding to $e^{-\tau h_{12}}$ in 1TS, this classification will change for other $e^{-\tau h_{ij}}$ at other time slices. For example, a path in $Z_\pm^{(0)}$ defined according to $e^{-\tau h_{12}}$ in 1TS may become one in $Z_\pm^{(1)}$ or $Z_\pm^{(2)}$ regarding to $e^{-\tau h_{ij}}$ at other time slices. However according to our definition of $Z_\pm$, paths in $Z_+^{(\alpha)}$ can only transfer into $Z_+^{(\alpha')}$, but not into $Z_-^{(\alpha'')}$; Similarly, paths in $Z_-^{(\alpha)}$ can only transfer into $Z_-^{(\alpha')}$, but not into $Z_+^{(\alpha'')}$. In the following discussion, unless otherwise specified, $Z_\pm^{(0)}$, $Z_\pm^{(1)}$ and $Z_\pm^{(2)}$ are defined for $e^{-\tau h_{12}}$ in 1TS.

In order to avoid mathematical ambiguous in the following derivation, $e^{-\tau h_{12}}$ in 1TS is formally replaced by $e^{-\tilde{\tau}h_{12}}$, and taking $\tilde{\tau}=\tau$ at the final step. Evidently,

$$Z_{+}^{(0)}=\sum_{K}\Big[\langle\overline{12}K|e^{-\beta H'}|\overline{12}K\rangle\langle\overline{12}K|e^{-\tilde{\tau}h_{ij}}|\overline{12}K\rangle_{+}$$

$$+\langle 12K|e^{-\beta H'}|12K\rangle\langle 12K|e^{-\tilde{\tau}h_{ij}}|12K\rangle_{+}\Big]$$

$$=\sum_{K}\Big[\langle\overline{12}K|e^{-\beta H'}|\overline{12}K\rangle_{+}+\langle 12K|e^{-\beta H'}|12K\rangle_{+}\Big]\qquad(4a)$$

$$Z_{-}^{(0)}=\sum_{K}\Big[\langle\overline{12}K|e^{-\beta H'}|\overline{12}K\rangle\langle\overline{12}K|e^{-\tilde{\tau}h_{ij}}|\overline{12}K\rangle_{-}$$

$$+\langle 12K|e^{-\beta H'}|12K\rangle\langle 12K|e^{-\tilde{\tau}h_{ij}}|12K\rangle_{-}\Big]$$

$$=\sum_{K}\Big[\langle\overline{12}K|e^{-\beta H'}|\overline{12}K\rangle_{-}+\langle 12K|e^{-\beta H'}|12K\rangle_{-}\Big]\qquad(4b)$$

$$Z_{+}^{(1)}=\sum_{K}\Big[\langle 1\overline{2}K|e^{-\beta H'}|1\overline{2}K\rangle\langle 1\overline{2}K|e^{-\tilde{\tau}h_{12}}|1\overline{2}K\rangle_{+}$$

$$+\langle\overline{1}2K|e^{-\beta H'}|\overline{1}2K\rangle\langle\overline{1}2K|e^{-\tilde{\tau}h_{12}}|\overline{1}2K\rangle_{+}\Big]$$

$$=\sum_{K}\Big[\langle 1\overline{2}K|e^{-\beta H'}|1\overline{2}K\rangle_{+}+\langle\overline{1}2K|e^{-\beta H'}|\overline{1}2K\rangle_{+}\Big]\frac{(e^{\tilde{\tau}t_{12}}+e^{-\tilde{\tau}t_{12}})}{2}\qquad(4c)$$

$$Z_{-}^{(1)}=\sum_{K}\Big[\langle 1\overline{2}K|e^{-\beta H'}|1\overline{2}K\rangle\langle 1\overline{2}K|e^{-\tilde{\tau}h_{12}}|1\overline{2}K\rangle_{-}$$

$$+\langle\overline{1}2K|e^{-\beta H'}|\overline{1}2K\rangle\langle\overline{1}2K|e^{-\tilde{\tau}h_{12}}|\overline{1}2K\rangle_{-}\Big]$$

$$=\sum_{K}\Big[\langle 1\overline{2}K|e^{-\beta H'}|1\overline{2}K\rangle_{-}+\langle\overline{1}2K|e^{-\beta H'}|\overline{1}2K\rangle_{-}\Big]\frac{(e^{\tilde{\tau}\theta_{12}t_{12}}+e^{-\tilde{\tau}\theta_{12}t_{12}})}{2}\qquad(4d)$$

$$Z_{+}^{(2)}=\sum_{K}\Big[\langle\overline{1}2K|e^{-\beta H'}|1\overline{2}K\rangle\langle 1\overline{2}K|e^{-\tilde{\tau}h_{12}}|\overline{1}2K\rangle_{+}$$

$$+\langle 1\overline{2}K|e^{-\beta H'}|\overline{1}2K\rangle\langle\overline{1}2K|e^{-\tilde{\tau}h_{12}}|1\overline{2}K\rangle_{+}\Big]$$

$$=\sum_{K}\Big[\langle\overline{1}2K|e^{-\beta H'}|1\overline{2}K\rangle_{+}+\langle 1\overline{2}K|e^{-\beta H'}|\overline{1}2K\rangle_{+}\Big]\frac{(e^{\tilde{\tau}t_{12}}-e^{-\tilde{\tau}t_{12}})}{2}\qquad(4e)$$

$$Z_{-}^{(2)} = \sum_{K} \Big[ \langle \bar{1}2K| e^{-\beta H'} |1\bar{2}K\rangle \langle 1\bar{2}K| e^{-\tilde{\tau} h_{12}} |\bar{1}2K\rangle_{-}$$

$$+ \langle 1\bar{2}K| e^{-\beta H'} |\bar{1}2K\rangle \langle \bar{1}2K| e^{-\tilde{\tau} h_{12}} |1\bar{2}K\rangle_{-} \Big]$$

$$= \sum_{K} \Big[ \langle \bar{1}2K| e^{-\beta H'} |1\bar{2}K\rangle_{-} + \langle 1\bar{2}K| e^{-\beta H'} |\bar{1}2K\rangle_{-} \Big] \frac{\left(e^{\tilde{\tau}\theta_{12} t_{12}} - e^{-\tilde{\tau}\theta_{12} t_{12}}\right)}{2} \quad (4f)$$

with $H' = H - \frac{\tilde{\tau}}{\beta} h_{12}$. The subscription $+$ and $-$ in above equations refer **OP** and **ON** paths, respectively. In the process of above deduction (Eq. 4a-4f), Eq. 2a-2c are used.

Considering $y_{\pm} = R\left(e^{\tilde{\tau}\theta t} \pm e^{-\tilde{\tau}\theta t}\right)$, where $\theta$ could be +1 or -1, and $R$ is independent on $\tilde{\tau}$. It is easily to prove,

$$-\frac{1}{y_{\pm}} \frac{\partial y_{\pm}}{\partial \tilde{\tau}} = -\theta t \frac{e^{\tilde{\tau}\theta t} \mp e^{-\tilde{\tau}\theta t}}{e^{\tilde{\tau}\theta t} \pm e^{-\tilde{\tau}\theta t}} = -t \frac{\left(e^{\tilde{\tau} t} \mp e^{-\tilde{\tau} t}\right)}{\left(e^{\tilde{\tau} t} \pm e^{-\tilde{\tau} t}\right)} \quad (5)$$

Using this equation, the contribution of $e^{-\tilde{\tau} h_{12}}$ in 1TS to energies reads,

$$\varepsilon_{12} = -\frac{1}{M} \frac{1}{Z} \frac{\partial Z}{\partial \tilde{\tau}}$$

$$= -\frac{t_{12}}{M} \frac{\left(Z_{+}^{(1)} + Z_{-}^{(1)}\right) \frac{e^{\tilde{\tau} t_{12}} - e^{-\tilde{\tau} t_{12}}}{e^{\tilde{\tau} t_{12}} + e^{-\tilde{\tau} t_{12}}} + \left(Z_{+}^{(2)} + Z_{-}^{(2)}\right) \frac{e^{\tilde{\tau} t_{12}} + e^{-\tilde{\tau} t_{12}}}{e^{\tilde{\tau} t_{12}} - e^{-\tilde{\tau} t_{12}}}}{Z_{+}^{(0)} + Z_{+}^{(1)} + Z_{+}^{(2)} + Z_{-}^{(0)} + Z_{-}^{(1)} + Z_{-}^{(2)}} \quad (6)$$

**Proof of** $\frac{Z_{-}^{(1)}}{Z_{+}^{(1)}} = \frac{Z_{-}^{(2)}}{Z_{+}^{(2)}}$**:** The Feynman path integral technique maps a D-dimensional quantum system to a (D+1)-dimensional classical system, in which a closed path in the quantum system corresponds a state of the classical system. If $\rho(\omega)$ is the weight of the path $\omega$, the effective energy ($E_{\omega}$) of this state in the corresponding classical system will be $E_{\omega} = -k_B T ln\rho(\omega)$. As the number of time-slices $M$ approaching to infinity, the size of the corresponding classical system also tends to the thermodynamic limit. When the size of the classical system is large enough, the thermal fluctuation can be neglected. In this condition, according to the statistical physics, almost all allowed paths are located at an isoenergetic surface, which corresponds to the most probable distribution. Namely, almost all allowed paths have the equal probability or weight. In fact, if the microcanonical ensemble is used, this point is guaranteed by the principle of equal probability or ergodic hypothesis. Given the fact that all OP paths have the positive global sign and all allowed paths have the equal probability, the function $f_+(\tilde{\tau}) = \frac{Z_+^{(1)}}{Z_+^{(2)}}$ is nothing but the ratio of the number of paths joining $Z_+^{(1)}$ over $Z_+^{(2)}$.

Next, we will prove the ratio $\frac{Z_+^{(2)}}{Z_+^{(1)}}\left(\frac{Z_-^{(2)}}{Z_-^{(1)}}\right)$ is uniquely determined by the bifurcation generated by $e^{-\tilde{\tau}h_{12}}$ in 1ST. For any WF, if one of site 1 and 2 is occupied, the scattering by $e^{-\tilde{\tau}h_{12}}$ will be bifurcated. One branch corresponds to the off-diagonal scattering (Eq. 2a), which contributes to $Z_\pm^{(2)}$. And another branch is the diagonal scattering (Eq. 2b) belonging to $Z_\pm^{(1)}$. We find that, except for those paths belonging to $Z_\pm^{(0)}$, the classification of a path is uncertain before scattered by $e^{-\tau h_{12}}$ in 1TS, which is illustrated in Fig. 1. As shown in Fig.1, an OP path starting at *A*, the classification of the path from *A* to *B* (the blue solid line in Fig. 1) is undefined. Once the path is scattered by $e^{-\tau h_{12}}$ in 1TS, the classification of the path will be determined. As shown in Fig. 1, after the scattering, the attribution of the path, either belonging to $Z_+^{(1)}$ (the red line in Fig. 1) or to $Z_+^{(2)}$ (the green line in Fig. 1) will be not changed by subsequent scatterings.

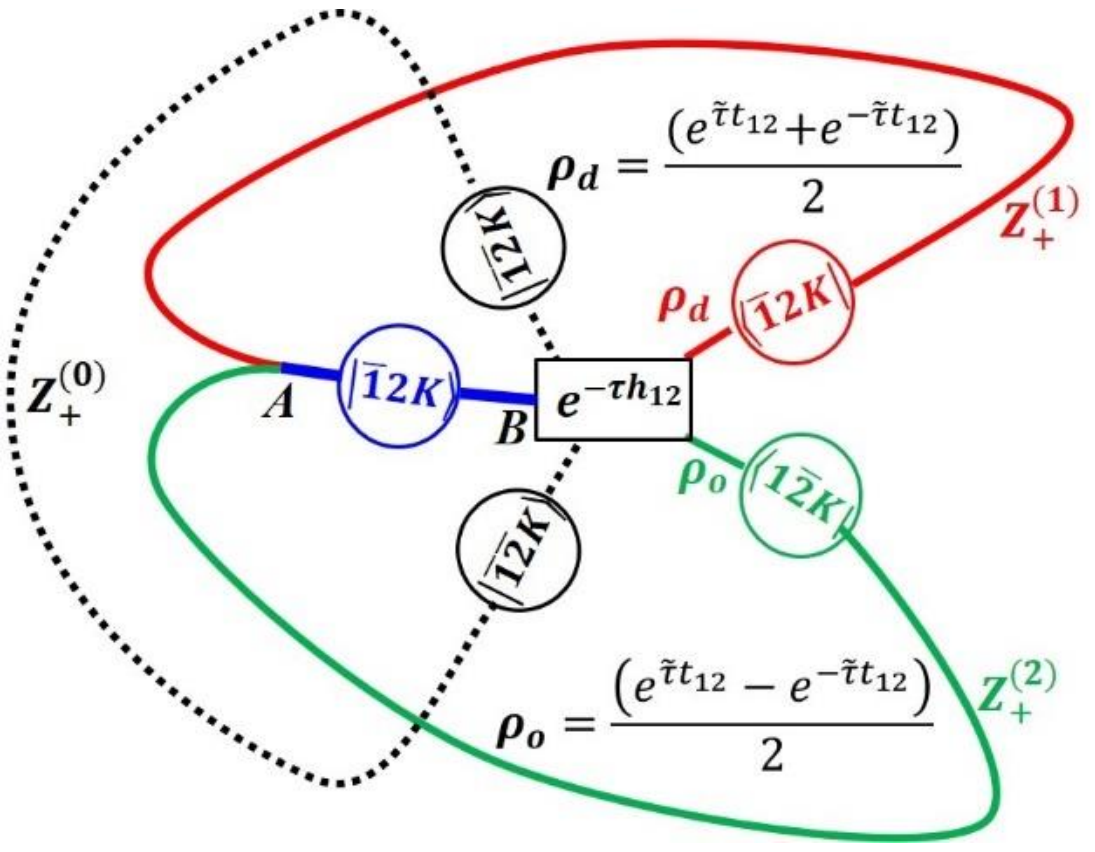

**Figure 1: Schematic illustration the classification of paths belonging to $Z_+^{(0)}$(black dot line), $Z_+^{(1)}$ (red solid line) or $Z_+^{(2)}$ (green solid line). This figure also shows how the scattering of $e^{-\tau h_{12}}$ in 1TS determining the classification of a path. Before the scattering happened, the classification of the path from *A* to *B* (the blue solid line) is undefined. Once the path is scattered by $e^{-\tau h_{12}}$ in 1TS, the**

**classification of the path, belongs to either $Z_+^{(1)}$ (the red line) or $Z_+^{(2)}$ (the green line), is fixed.**

According to above discussion, $f_+(\tilde{\tau})$ should be completely determined by the scattering of $e^{-\tau h_{12}}$ in 1TS. Physically $f_+(\tilde{\tau})$ consists of two factors: the local BW determining the choice of one path between $Z_+^{(1)}$ and $Z_+^{(2)}$ and the number of available channels to $Z_+^{(1)}$ and $Z_+^{(2)}$. Define $f_+(\tilde{\tau}) = p_+ r_+$. $p_+$ refers the ratio of local BWs between two branches of a bifurcation. And $r_+$ is the ratio of the number of channels between $Z_+^{(1)}$ and $Z_+^{(2)}$. As shown in Fig. 1, in one branch (red line), $e^{-\tau h_{12}}$ in 1TS takes the diagonal scattering, which has the weight of $\frac{(e^{\tilde{\tau} t_{12}} + e^{-\tilde{\tau} t_{12}})}{2}$. While in another branch (green line), $e^{-\tau h_{12}}$ in 1TS takes the off-diagonal scattering, which has the weight of $\frac{|e^{\tilde{\tau} t_{12}} - e^{-\tilde{\tau} t_{12}}|}{2}$. Namely $p_+ = \frac{(e^{\tilde{\tau} t_{12}} + e^{-\tilde{\tau} t_{12}})}{|e^{\tilde{\tau} t_{12}} - e^{-\tilde{\tau} t_{12}}|}$.[77] $\frac{(e^{\tilde{\tau} t_{12}} + e^{-\tilde{\tau} t_{12}})}{2}$ is always larger than zero, but $\frac{(e^{\tilde{\tau} t_{12}} - e^{-\tilde{\tau} t_{12}})}{2}$ may be positive or negative depending on the sign of $t_{12}$ and $\theta_{12}$. To make $p_+ = \frac{(e^{\tilde{\tau} t_{12}} + e^{-\tilde{\tau} t_{12}})}{|e^{\tilde{\tau} t_{12}} - e^{-\tilde{\tau} t_{12}}|}$ hold, the sign of $\frac{(e^{\tilde{\tau} t_{12}} - e^{-\tilde{\tau} t_{12}})}{2}$ must be the same in each scattering by $e^{-\tau h_{12}}$ in 1TS. This is why an OP path requires the positive local sign ($\theta_{12} = 1$). Obviously, our definition of OP paths is made to meet this condition. Another crucial feature of our classification ($Z_+^{(1,2,3)}$) is that, the two branches from the same bifurcation must belong to the OP space simultaneously. And it is impossible for one branch in ON space and the other branch in OP space. Namely the paths in the OP (ON) space form a **self-closed zone**. The self-closed character indicates that the number of channels to $Z_+^{(1)}$ and to $Z_+^{(2)}$ is exactly the same, thus $r_+ = 1$.

Similarly, define $f_-(\tilde{\tau}) = \frac{Z_-^{(1)}}{Z_-^{(2)}}$. The above procedure can be extended to ON paths, and the proof of $f_-(\tilde{\tau}) = \frac{(e^{\tilde{\tau}\theta_{12} t_{12}} + e^{-\tilde{\tau}\theta_{12} t_{12}})}{|e^{\tilde{\tau}\theta_{12} t_{12}} - e^{-\tilde{\tau}\theta_{12} t_{12}}|}$ is straightforwardly. Only one point needs to be stressed that, to prove $f_-(\tilde{\tau}) = \frac{(e^{\tilde{\tau}\theta_{12} t_{12}} + e^{-\tilde{\tau}\theta_{12} t_{12}})}{|e^{\tilde{\tau}\theta_{12} t_{12}} - e^{-\tilde{\tau}\theta_{12} t_{12}}|}$, the paths in $Z_-^{(1)}$ and $Z_-^{(2)}$ need to be further divided into four categories. The first class ($Q_1^{(\alpha)}$, $\alpha = 1,2$) has the negative global sign and $\theta_{12}=1$; The second one ($Q_2^{(\alpha)}$, $\alpha=1{,}2$) has the negative global sign and $\theta_{12}=-1$; The third one ($Q_3^{(\alpha)}$, $\alpha=1{,}2$) has the positive global sign and $\theta_{12}=1$; The fourth one ($Q_4^{(\alpha)}$, $\alpha=1{,}2$) has the positive global sign and $\theta_{12}=-1$. Obviously, $Z_-^{(\alpha)}=Q_1^{(\alpha)}+Q_2^{(\alpha)}+Q_3^{(\alpha)}+Q_4^{(\alpha)}$. It is easy to see that, the above classification has three features: 1) In each classification, the global sign of all paths is the same; 2) In each classification, the local sign $\theta_{12}$ of all paths is the same, which guarantees $\frac{(e^{\tilde{\tau}\theta_{12}t_{12}}-e^{-\tilde{\tau}\theta_{12}t_{12}})}{2}$ having the same sign. 3) There is the self-closed character in each classification. As we discussed above, the three features are the essential precondition of the proof. Following the similar steps in proving $f_+(\tilde{\tau})$, we can easily prove $\frac{Q_\vartheta^{(1)}}{Q_\vartheta^{(2)}}=\frac{(e^{\tilde{\tau}\theta_{12}t_{12}}+e^{-\tilde{\tau}\theta_{12}t_{12}})}{|e^{\tilde{\tau}\theta_{12}t_{12}}-e^{-\tilde{\tau}\theta_{12}t_{12}}|}$ for $\vartheta=1,2,3$ or 4. Accordingly, we have $f_-(\tilde{\tau})=\frac{(e^{\tilde{\tau}\theta_{12}t_{12}}+e^{-\tilde{\tau}\theta_{12}t_{12}})}{|e^{\tilde{\tau}\theta_{12}t_{12}}-e^{-\tilde{\tau}\theta_{12}t_{12}}|}$ too.

We finally obtain, $\frac{Z_+^{(1)}}{Z_+^{(2)}}=f_+(\tilde{\tau})=f_-(\tilde{\tau})=\frac{Z_-^{(1)}}{Z_-^{(2)}}$. By reforming this equation, we have,

$$\frac{Z_-^{(1)}}{Z_+^{(1)}}=\frac{Z_-^{(2)}}{Z_+^{(2)}}=\gamma \qquad (7)$$

Note, $\gamma$ is another function different from $f_\pm(\tilde{\tau})$.

**Proof of $\frac{Z_-^{(0)}}{Z_+^{(0)}}=\frac{Z_-^{(1)}}{Z_+^{(1)}}=\frac{Z_-^{(2)}}{Z_+^{(2)}}$:** In the subset path space of $Z^{(1+2)}=Z_+^{(1)}+Z_+^{(2)}+Z_-^{(1)}+Z_-^{(2)}$, the expectation value of $h_{12}$ at 1TS can be calculated according to the thermodynamic estimator of energies,

$$\begin{aligned}\langle h_{12}\rangle_{Z^{(1+2)}} &= -\frac{1}{M}\frac{1}{Z^{(1+2)}}\frac{\partial Z^{(1+2)}}{\partial\tilde{\tau}}\\
&= -\frac{t_{12}}{M}\frac{(Z_+^{(1)}+Z_-^{(1)})\frac{e^{\tilde{\tau}t_{12}}-e^{-\tilde{\tau}t_{12}}}{e^{\tilde{\tau}t_{12}}+e^{-\tilde{\tau}t_{12}}}+(Z_+^{(2)}+Z_-^{(2)})\frac{e^{\tilde{\tau}t_{12}}+e^{-\tilde{\tau}t_{12}}}{e^{\tilde{\tau}t_{12}}-e^{-\tilde{\tau}t_{12}}}}{Z_+^{(1)}+Z_+^{(2)}+Z_-^{(1)}+Z_-^{(2)}}\\
&= -\frac{t_{12}}{M}\frac{Z_+^{(1)}(1+\gamma)\frac{e^{\tilde{\tau}t_{12}}-e^{-\tilde{\tau}t_{12}}}{e^{\tilde{\tau}t_{12}}+e^{-\tilde{\tau}t_{12}}}+Z_+^{(2)}(1+\gamma)\frac{e^{\tilde{\tau}t_{12}}+e^{-\tilde{\tau}t_{12}}}{e^{\tilde{\tau}t_{12}}-e^{-\tilde{\tau}t_{12}}}}{Z_+^{(1)}(1+\gamma)+Z_+^{(2)}(1+\gamma)}\\
&= -\frac{t_{12}}{M}\frac{Z_+^{(1)}\frac{e^{\tilde{\tau}t_{12}}-e^{-\tilde{\tau}t_{12}}}{e^{\tilde{\tau}t_{12}}+e^{-\tilde{\tau}t_{12}}}+Z_+^{(2)}\frac{e^{\tilde{\tau}t_{12}}+e^{-\tilde{\tau}t_{12}}}{e^{\tilde{\tau}t_{12}}-e^{-\tilde{\tau}t_{12}}}}{Z_+^{(1)}+Z_+^{(2)}} \qquad (8a)\end{aligned}$$

In deriving Eq. 8a, Eq. 7 is used. According to $Z^{(1+2)} = (1+\gamma)(Z_+^{(1)} + Z_+^{(2)})$, $\langle h_{12} \rangle_{Z^{(1+2)}}$ can also be calculated as,

$$\langle h_{12} \rangle_{Z^{(1+2)}} = -\frac{1}{M}\frac{1}{Z^{(1+2)}}\frac{\partial Z^{(1+2)}}{\partial \tilde{\tau}}$$

$$= -\frac{t_{12}}{M}\frac{Z_+^{(1)}\frac{e^{\tilde{\tau}t_{12}} - e^{-\tilde{\tau}t_{12}}}{e^{\tilde{\tau}t_{12}} + e^{-\tilde{\tau}t_{12}}} + Z_+^{(2)}\frac{e^{\tilde{\tau}t_{12}} + e^{-\tilde{\tau}t_{12}}}{e^{\tilde{\tau}t_{12}} - e^{-\tilde{\tau}t_{12}}}}{\left(Z_+^{(1)} + Z_+^{(2)}\right)} - \frac{t_{12}}{M}\frac{\partial \gamma}{(1+\gamma)\partial \tilde{\tau}} \quad (8b)$$

$\langle h_{12} \rangle_{Z^{(1+2)}}$ obtained in Eq. 8a and 8b must be identical, which implies $\frac{\partial \gamma}{\partial \tilde{\tau}} = 0$, namely $\gamma$ is independent on temperature!

Recall $h_{12} = -t_{12}(c_1^\dagger c_2 + c_1^\dagger c_2)$, within the subset path space of $Z^{(1+2)}$, the expectation value of $c_1^\dagger c_2 + c_1^\dagger c_2$ at 1TS can be calculated as,

$$\langle c_i^\dagger c_j + c_j^\dagger c_i \rangle_{Z^{(1+2)}} = \frac{1}{Z^{(1+2)}}\frac{\partial Z^{(1+2)}}{\partial (\tilde{\tau}t_{12})}$$

$$= \frac{(Z_+^{(1)} + Z_-^{(1)})\frac{e^{\tilde{\tau}t_{12}} - e^{-\tilde{\tau}t_{12}}}{e^{\tilde{\tau}t_{12}} + e^{-\tilde{\tau}t_{12}}} + (Z_+^{(2)} + Z_-^{(2)})\frac{e^{\tilde{\tau}t_{12}} + e^{-\tilde{\tau}t_{12}}}{e^{\tilde{\tau}t_{12}} - e^{-\tilde{\tau}t_{12}}}}{Z_+^{(1)} + Z_+^{(2)} + Z_-^{(1)} + Z_-^{(2)}}$$

$$= \frac{Z_+^{(1)}(1+\gamma)\frac{e^{\tilde{\tau}t_{12}} - e^{-\tilde{\tau}t_{12}}}{e^{\tilde{\tau}t_{12}} + e^{-\tilde{\tau}t_{12}}} + Z_+^{(2)}(1+\gamma)\frac{e^{\tilde{\tau}t_{12}} + e^{-\tilde{\tau}t_{12}}}{e^{\tilde{\tau}t_{12}} - e^{-\tilde{\tau}t_{12}}}}{Z_+^{(1)}(1+\gamma) + Z_+^{(2)}(1+\gamma)}$$

$$= \frac{Z_+^{(1)}\frac{e^{\tilde{\tau}t_{12}} - e^{-\tilde{\tau}t_{12}}}{e^{\tilde{\tau}t_{12}} + e^{-\tilde{\tau}t_{12}}} + Z_+^{(2)}\frac{e^{\tilde{\tau}t_{12}} + e^{-\tilde{\tau}t_{12}}}{e^{\tilde{\tau}t_{12}} - e^{-\tilde{\tau}t_{12}}}}{Z_+^{(1)} + Z_+^{(2)}} \quad (9a)$$

Similarly, according to $Z^{(1+2)} = (1+\gamma)(Z_+^{(1)} + Z_+^{(2)})$, we also have,

$$\langle c_i^\dagger c_j + c_j^\dagger c_i \rangle_{Z^{(1+2)}} = \frac{1}{Z^{(1+2)}}\frac{\partial Z^{(1+2)}}{\partial (\tilde{\tau}t_{12})}$$

$$= \frac{Z_+^{(1)}\frac{e^{\tilde{\tau}t_{12}} - e^{-\tilde{\tau}t_{12}}}{e^{\tilde{\tau}t_{12}} + e^{-\tilde{\tau}t_{12}}} + Z_+^{(2)}\frac{e^{\tilde{\tau}t_{12}} + e^{-\tilde{\tau}t_{12}}}{e^{\tilde{\tau}t_{12}} - e^{-\tilde{\tau}t_{12}}}}{\left(Z_+^{(1)} + Z_+^{(2)}\right)} + \frac{\partial \gamma}{(1+\gamma)\partial (\tilde{\tau}t_{12})} \quad (9b)$$

$\langle c_i^\dagger c_j + c_j^\dagger c_i \rangle_{Z^{(1+2)}}$ obtained in Eq. 9a and 9b must be identical too, thus $\gamma$ is also independent on the hopping amplitude. Since $e^{-\tau h_{12}}$ in 1TS is nothing special, the above proof equation can be extended to any type of $h_{ij}$ at any time slice. Finally, we reach the essential conclusion that $\gamma$ is independent on both $\tau$ and $t_{ij}$ for any $e^{-\tau h_{ij}}$ at any time slice. In other words, $\gamma$ is independent on both temperature and interaction.

Define $\sigma = \frac{Z_-}{Z_+} = \frac{Z_-^{(0)}+Z_-^{(1)}+Z_-^{(2)}}{Z_+^{(0)}+Z_+^{(1)}+Z_+^{(2)}}$, we have,

$$\sigma = \gamma + \frac{(\vartheta - \gamma)Z_+^{(0)}}{Z_+^{(0)} + Z_+^{(1)} + Z_+^{(2)}}. \quad (10)$$

with $\vartheta = \frac{Z_-^{(0)}}{Z_+^{(0)}}$. It is self-evident that, no matter how the classification changes, both $Z_-$ and $Z_+$, as well as the ratio of $Z_-$ to $Z_+$, must have a definite value for a given system and temperature. For any $e^{-\tau h_{ij}}$ at any time slice, and whatever $h_{ij}$ changing with $i$ and $j$, Eq. 10 must hold. However, $Z_+^{(0)}$ is changeable with $e^{-\tau h_{ij}}$ and time slice. Since $Z_+^{(0)} + Z_+^{(1)} + Z_+^{(2)}$ is also a constant at a given temperature, to guarantee $\sigma$ unchanged, the only possibility is $\vartheta = \gamma$, namely,

$$\frac{Z_-^{(0)}}{Z_+^{(0)}} = \frac{Z_-^{(1)}}{Z_+^{(1)}} = \frac{Z_-^{(2)}}{Z_+^{(2)}} = \gamma \qquad (11)$$

Using the proportional relationship of Eq.11, Eq. 6 can be written as,

$$\varepsilon_{12} = -\frac{t_{12}}{M}\frac{Z_+^{(1)}\frac{e^{-\tilde{\tau}t_{12}}-e^{\tilde{\tau}t_{12}}}{e^{-\tilde{\tau}t_{12}}+e^{\tilde{\tau}t_{12}}}+Z_+^{(2)}\frac{e^{-\tilde{\tau}t_{12}}+e^{\tilde{\tau}t_{12}}}{e^{-\tilde{\tau}t_{12}}-e^{\tilde{\tau}t_{12}}}}{Z_+^{(0)}+Z_+^{(1)}+Z_+^{(2)}} = -\frac{t_{12}}{M}\frac{\partial Z_+}{Z_+\partial\tilde{\tau}}$$

Since $h_{12}$ in 1TS is nothing special, the above equation can be used to calculate expectation value of $h_{ij}$ at any time slice in OP path space. Although the above analysis is focus on energies, other physics quantities can be analyzed in the similar way, and calculated only in OP paths. If $\hat{O}$ is a physics quantity, $\langle\hat{O}\rangle$ reads

$$\langle\hat{O}\rangle = \langle\hat{O}\rangle_{OP} = \left(\frac{Tr\hat{O}e^{-\beta H}}{Tre^{-\beta H}}\right)_{OP} \qquad (12)$$

Since all OP paths always have the global positive sign, thus SP completely disappears!

**General Remarks:** The following issues need to be stressed. **1)** The key step in the current proof lies on the combination of each off-diagonal term in Hamiltonian and its Hermitian conjugate. Without this combination, the strict mathematical proof is definitely impossible. **2)** In above analysis, an implicit feature of $\frac{\partial Z_+^{(\alpha)}}{Z_+^{(\alpha)}\partial\tilde{\tau}} = \frac{\partial Z_-^{(\alpha)}}{Z_-^{(\alpha)}\partial\tilde{\tau}}$, namely OP and ON paths contributing the identical instantaneous energy, is crucial, especially in the mathematical derivation of Eq. 7-10. With this feature, we know that the cancellation only modifies the weight of paths, but does not change the instantaneous local energy. This is the reason why we work on the local Boltzmann operator ($e^{-\tau h_{ij}}$) rather than the global Boltzmann operator ($e^{-\beta H}$).[84] **3)** Although the above proof is for one-body operators, it is straightforward to extend to two-body and even more complex operators, and the conclusion will not change. **4)** The abandoned paths that belong $Z_-$ (or ON paths) can have either positive- or negative-weight. Clearly, we do not simply abandon negative-weighted paths. **5)** The above discussion does not involve Hubbard-Stratonovich transformation, however the current method can be straightforwardly adopted to the effective Hamiltonian introduced by the Hubbard-Stratonovich transformation. And the current idea for solving the SP is independent on the specified QMC algorithm, the only requirement is to combine each off-diagonal term in Hamiltonian and its Hermitian conjugate into pairs, and constrains the sampling of QMC within the OP paths. **6)**, The WF space contains a huge number of individual paths, a path can be defined as an individual path or a group of a few individual paths. In different QMCs, the meaning of a path may be quite different. In fact, a path defined in current work is different from that used in other QMCs. The most essential point is that, our strategy for SP and our path integral formula are inseparable, which should be considered as a whole. **Finally,** our proof is strict and does not involve any approximation so far, which is completely different from those based on fermion nodes of density matrix. It has been well known that, if the nodes of the many-fermion density matrix were known, the properties of fermion systems could be exactly calculated by QMC methods, and at the same time the SP was avoided.[16] However, for almost all realistic systems the nodes are unknown, thus the method based on fermion nodes must be included approximations. On the contrary, OP paths suggested in the current work can be easily constructed, and it is not necessary to know the boundary of OP spaces. Additionally, by combining configuration interaction (CI) and QMC methods, some approaches in use today do alleviate the SP remarkably.[85] However, those approaches are still an approximation method, and need the heavy computational effort.

## III. Practical Consideration and Strategy

Although Eq. 12 does avoid the SP, an efficient QMC method targeted at the current idea needs to be developed. Here we try to implement this idea into our recent proposed world-line-based QMC method.[77] We find that the difficult lies in how to sample a closed path in the OP space. Since all local signs in an OP path must be positive, it imposes a very strong requirement for the occupation state in WFs, which is very time-consuming to generate a closed path in our world-line-based QMC method. In order to overcome this difficulty, we try to open a small window, from which the positive and negative local sign can coexist. To this purpose, we redivide the closed paths into two classes: **RP** and **RN**. A RP path has the positive global sign, and all its off-diagonal scatterings have the positive local sign, *i.e*, $\theta_{ij}$=1 in Eq. 2a. All rest paths belong to RN. An RN path either has a global negative sign or contains at least one negative off-diagonal scattering. The difference between RP and OP paths lies in the local sign of diagonal scatterings (Eq. 2b). **For OP paths, the local sign of diagonal scatterings can only be positive, while it can be both positive and negative in RP paths.** Similarly, the partition function can be divided into three parts, $Z_+ = \tilde{Z}_+^{(0)} + \tilde{Z}_+^{(1)} + \tilde{Z}_+^{(2)}$ and $Z_- = \tilde{Z}_-^{(0)} + \tilde{Z}_-^{(1)} + \tilde{Z}_-^{(2)}$. Hereafter the subscript $+$ and $-$ refer the RP and RN space, respectively. In the following, we will demonstrate the QMC simulation can be carried out at the RP space instead of the OP space.

The mathematical proofs in the last section require paths in OP form a self-closed zone, so do that in ON. However, paths in RP or RN have not the self-closed character. As illustrated in Fig. 2a, consider a path in the RP space, $s^{(1)} = \left\{\langle 1\bar{2}K|e^{-\beta H'}|1\bar{2}K\rangle_+ + \langle \bar{1}2K|e^{-\beta H'}|\bar{1}2K\rangle_+\right\}\frac{(e^{\tilde{\tau}\theta_{12}t_{12}}+e^{-\tilde{\tau}\theta_{12}t_{12}})}{2}$, which belongs to $\tilde{Z}_+^{(1)}$. For $\theta_{12} = 1$, the scattering path of $s^{(1)}$ reads $s^{(2)} = \left\{\langle 1\bar{2}K|e^{-\beta H'}|1\bar{2}K\rangle_+ + \langle \bar{1}2K|e^{-\beta H'}|\bar{1}2K\rangle_+\right\}\frac{(e^{\tilde{\tau}t_{12}}-e^{-\tilde{\tau}t_{12}})}{2}$, which belong to $\tilde{Z}_+^{(2)}$ and still lays in the RP space. However, for $\theta_{12} = -1$, the scattering path of $s^{(1)}$, $s^{(2)} = \left\{\langle 1\bar{2}K|e^{-\beta H'}|1\bar{2}K\rangle_+ + \langle \bar{1}2K|e^{-\beta H'}|\bar{1}2K\rangle_+\right\}\frac{(e^{-\tilde{\tau}t_{12}}-e^{\tilde{\tau}t_{12}})}{2}$ belongs to $\tilde{Z}_-^{(2)}$ and does not lay in the RP space

any more (see Fig. 2b).

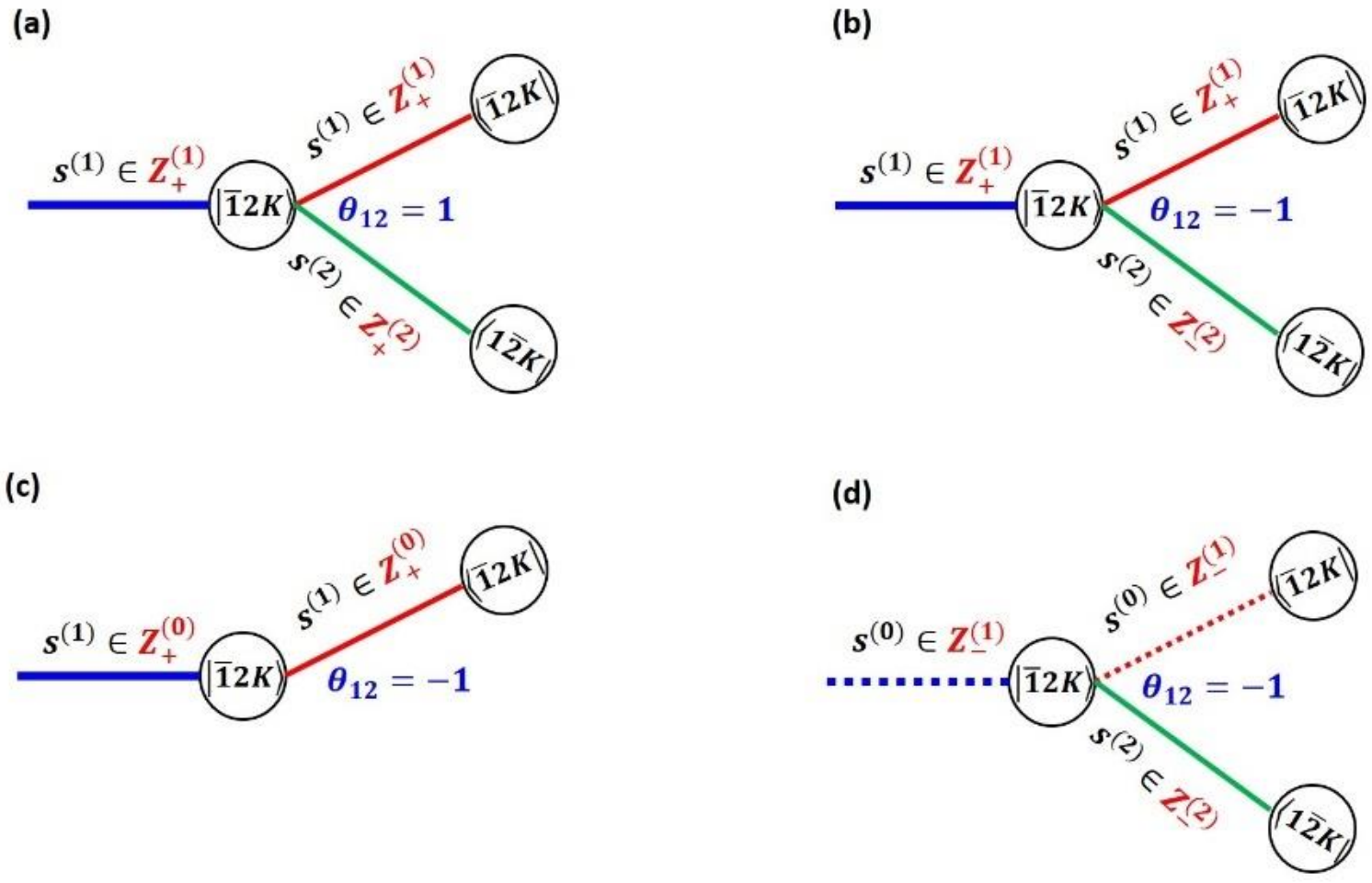

**Figure 2: Schematic illustration how to repair the self-closed character in the RP and RN space. (a) Both $s^{(1)}$ and its scattering path $s^{(2)}$ belong RP space for $\theta_{12} = 1$. (b) For $\theta_{12} = -1$, $s^{(1)}$ and its scattering path $s^{(2)}$ belong RP and RN space, respectively, which results in the breakdown of the self-closed character. (c) In RP space, all paths like $s^{(1)}$ with $\theta_{12} = -1$ are entirely absorbed into $\tilde{Z}_{+}^{(0)}$, thus the self-closed character is restored. (d) For $\theta_{12} = -1$, some paths from $\tilde{Z}_{-}^{(0)}$ are moved to $\tilde{Z}_{-}^{(1)}$, which form the scattering pairs with $s^{(2)}$, thus the self-closed character is also restored in RN space.**

In order to use the above mathematical proof, the self-closed character in the RP and RN space must be repaired. We find that, as $\tilde{\tau} \to 0$, the contribution of $\tilde{Z}_{+}^{(1)}$ to energies approaches to zero, which is much close the that of $\tilde{Z}_{+}^{(0)}$ and is much smaller than that of $\tilde{Z}_{+}^{(2)}$ (see Eq. 6). In order to restore the self-closed character in RP space, all paths like $s^{(1)}$ with $\theta_{12} = -1$ are entirely absorbed into $\tilde{Z}_{+}^{(0)}$. This means that formally these paths will not be scattered in the RP space, which is illustrated in Fig. 2c. With this operation, paths in RP form a self-closed zone again. Similarly, in the RN space, the opposite process is taken, as shown in Fig. 2d. Specifically, some paths in $\tilde{Z}_{-}^{(0)}$ are transferred into $\tilde{Z}_{-}^{(1)}$. These transferred paths are used to form scattering pairs with $s^{(2)}$, which is originally scattered from RP space. After this re-classification, paths in RN space form a self-closed zone too. Once the self-closed character is reestablished in both RP and RN independently, the mathematical proof in the last section can be copied here verbatim. Finally, we achieve the following key equation,

$$\langle\hat{O}\rangle \cong \langle\hat{O}\rangle_{RP} = \left(\frac{Tr\hat{O}e^{-\beta H}}{Tre^{-\beta H}}\right)_{RP} \qquad (13)$$

Once the self-closed character is restored in both RP and RN spaces, $r_+ = r_- = 1$ will automatically fulfilled. However, after some paths in $\tilde{Z}_+^{(0)}(\tilde{Z}_-^{(0)})$ are exchanged with that in $\tilde{Z}_+^{(1)}(\tilde{Z}_-^{(1)})$, $p_+ = p_-$ will no longer be exact but an approximation. The error caused by this exchange is in the order of $\tau^2$ on average, which can be seen in the following. When a path from $\tilde{Z}_\pm^{(0)}$ moves to $\tilde{Z}_\pm^{(1)}$, and vice versa, the probability ratio between the *fake* scattering pairs will be $\frac{2}{|e^{\tilde{\tau}t_{12}}-e^{-\tilde{\tau}t_{12}}|}$ rather than $\frac{(e^{\tilde{\tau}t_{12}}+e^{-\tilde{\tau}t_{12}})}{|e^{\tilde{\tau}t_{12}}-e^{-\tilde{\tau}t_{12}}|}$. In both RP and RN spaces, $p_\pm$ should be a number between $\frac{2}{|e^{\tilde{\tau}t_{12}}-e^{-\tilde{\tau}t_{12}}|}$ and $\frac{(e^{\tilde{\tau}t_{12}}+e^{-\tilde{\tau}t_{12}})}{|e^{\tilde{\tau}t_{12}}-e^{-\tilde{\tau}t_{12}}|}$ in general. Thus, the error in $p_\pm$ is roughly $\frac{2}{(e^{\tilde{\tau}t_{12}}+e^{-\tilde{\tau}t_{12}})}$, namely in the order of $\tau^2$. In fact, there are the huge number of paths in both $\tilde{Z}_\pm^{(0)}$ and $\tilde{Z}_\pm^{(1)}$. If the paths exchanging between $\tilde{Z}_\pm^{(0)}$ and $\tilde{Z}_\pm^{(1)}$ follows a specific rule, this approximation can be much improved. The rule is to choose those paths so that $p_\pm$ being as close to $\frac{(e^{\tilde{\tau}t_{12}}+e^{-\tilde{\tau}t_{12}})}{|e^{\tilde{\tau}t_{12}}-e^{-\tilde{\tau}t_{12}}|}$ as possible. It needs to point out that, the path exchange is just for mathematic proof not for QMC simulation. Our calculations in this paper are based on Eq. 13, and the results manifest that Eq.13 is an excellent approximation of Eq. 12.

It should be stressed in particular that, Eq. 12 is mathematically rigorous. At present, using Eq. 13 instead of Eq. 12 is mainly due to the fact that we have not developed a set of methods to efficiently sample paths in OP space. However, our current QMC code works well in RP space.

## IV. Apply to Hubbard Model

The QMC simulation is based on the world-line algorithm recently proposed in our group.[77] In this method, to obtain a closed path (a non-zero weight path), we design an algorithm similar to the world-line algorithm[86] and the multiple time threading one.[3] If without any constraints, each path may have the different global sign and local sign, namely it can be RP or RN paths. In order to implement the current idea, or specifically, in order to calculate Eq.13, we need to make appropriate modifications to the QMC algorithm. In fact, this modification is very simple. We only need to make QMC calculations in RP paths. Specifically, we assign zero weight to each RN path, accordingly QMC simulation will be naturally constrained in RP paths.

The physics properties of the two-dimensional Hubbard model with the size of $4 \times 4$ are calculated at finite temperature. In all calculations, $t_{ij} = t$ is taken as the unit of energy and set $t$ = 1.0 and $U = 4$. We focus on the system with 14 electrons, $i.e., N = 14$, corresponding to the electron density $\rho =$ 0.875. And the number of spin-up and spin-down electrons is equal. The system with $\rho = 0.875$ has the heaviest SP in the standard Hubbard model.[67] The total number of QMC steps at each temperature or $M$ is in the range of [$10^7$, $10^{10}$], where the first $3\times10^6$ steps are used to equilibrate the system and the rest steps are used to calculate the physical properties. According to thermodynamics, the energy is calculated according to $E = -\frac{1}{Z} \cdot \frac{\partial Z}{\partial \beta}$, and the double occupancy are calculated as $O_d = \sum_i < n_{i\uparrow} n_{i\downarrow} >$. The error bar is estimated by the so-called renormalization group method, which is designed for treating correlated data.[87]

The convergence tests of $\tau$ for the system at the temperature of 0.5 are shown in Fig. 3. It can be seen that, with the increase of $M$, namely $\tau$ being decreased, the energy is quickly convergent. From $M$ = 40, *i.e.*, $\tau$ = 0.05, the change in QMC results is much slowly, indicating the convergence. In the rest QMC simulations, $\tau$ is fixed at the value of 0.05.

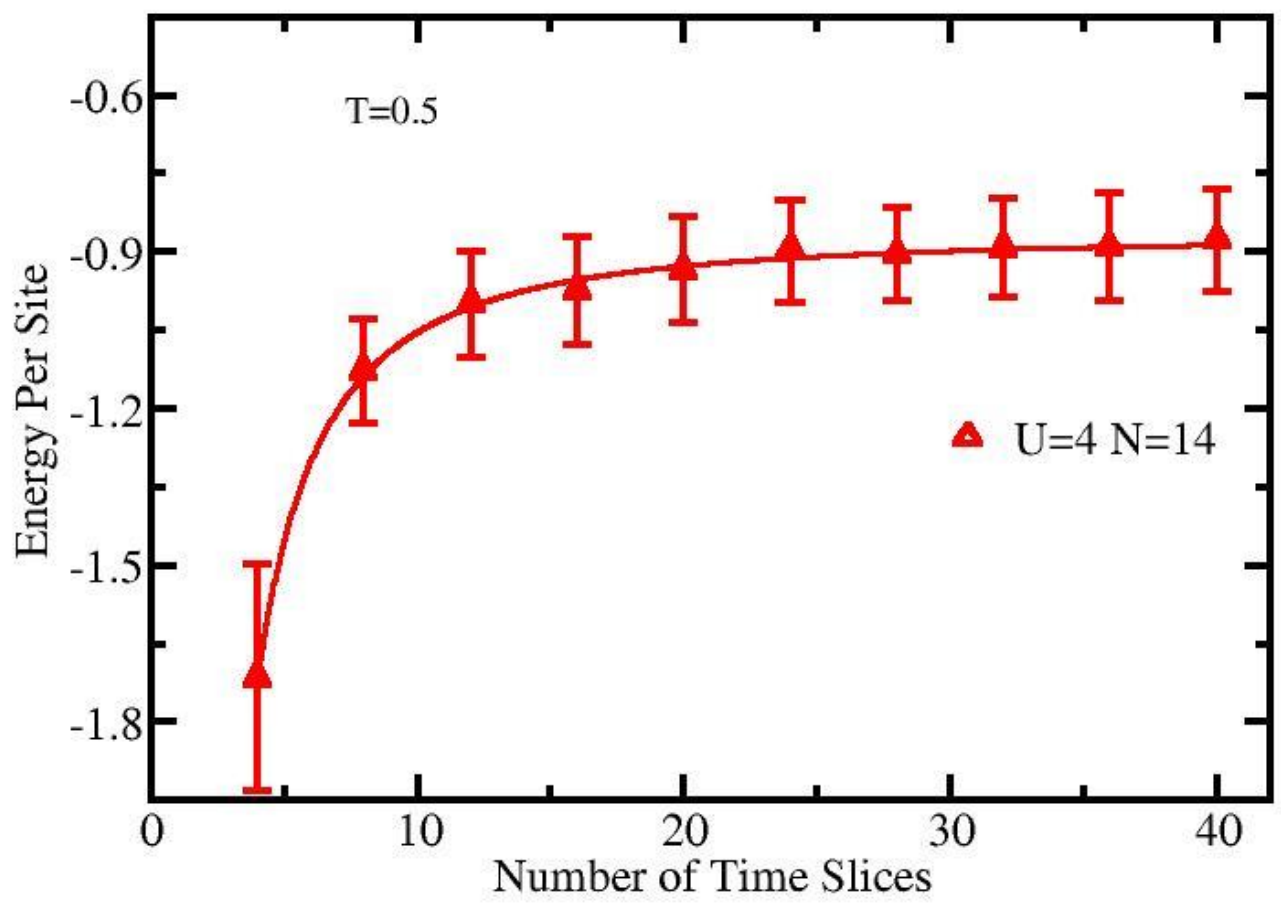

**Figure 3: The energy with the number of time slices (*M*) at the temperature of 0.5 for $N = 14$ and $U = 4.0$ system. The QMC calculation becomes convergent for $M \geq 40$.**

Since the current QMC is limited in RP paths, the SP is automatically absent. In the following, we will compare the current results to that presented in literatures. We find that, the energy and double occupancy calculated by current QMC simulations are in excellent agreement with that from various techniques[79] in the entire range of temperatures. Fig. 4 depicts the energy (upper panel) and double occupancy (lower panel) via temperature for $N = 14$ and $U = 4.0$. From this figure, one can see that, the current results for both the energy and double occupancy are in excellent agreement with previous data. In this figure, the data from Ref.[79] is calculated by the dynamical cluster approximation (DCA) and bare and bold-line diagrammatic Monte Carlo (DiagMC) method. The data labeled as "DCA_inf" and "DiagMC" represents the extrapolation of results to the thermodynamic limit. While the label "DCA_*xx*" is the finite results with the system size of *xx* (*xx*=20,50,32, and 34).[79]

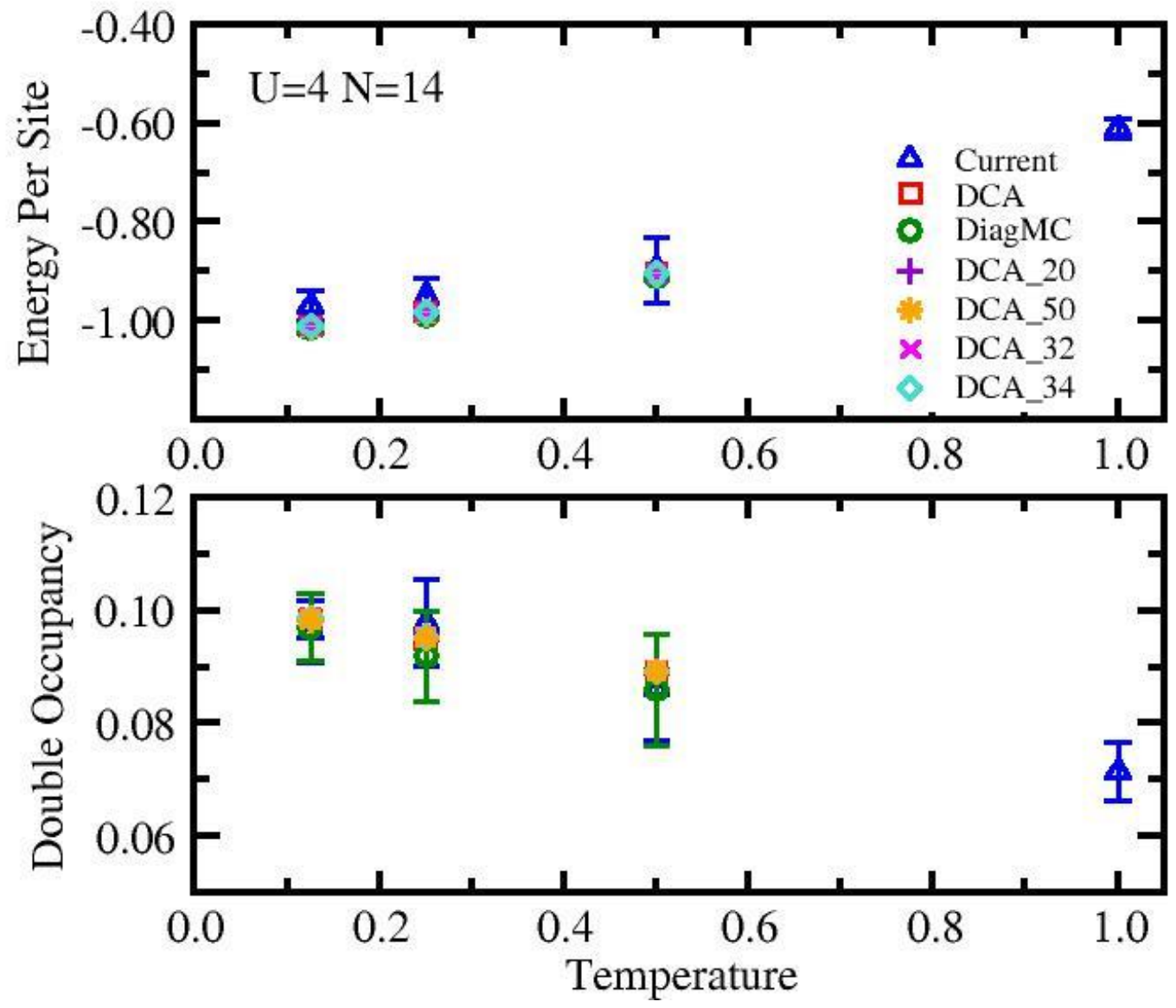

**Figure 4: Temperature dependence on the energy (upper panel) and double occupancy (lower panel) for $N = 14$ and $U = 4.0$. The triangles represent the current results, while the rest symbols represent the results from various techniques.[79]**

## V. Summary

In summary, we have found a rigorous solution for the fermion sign problem. The basic idea behind this solution is to look for a way to completely cancel or replace those paths with negative weights. By analyzing the contribution of a local Boltzmann operator to the partition function, we have proved that, the physical quantities can be exactly calculated some specific positive-weighted paths, thus the fermion sign problem is completely and rigorously solved. As an example, we have calculated the thermodynamic quantities of two-dimensional Hubbard model at finite temperature. Our results are in excellent agreement with previous values, which confirms the reliability of the current method.

**Acknowledgements:** Project supported by the National Natural Science Foundation of China (Grant No. 11874148). The computations were supported by ECNU Public Platform for Innovation.